\documentclass[envcountsame,11pt,a4paper]{llncs}

\usepackage{a4wide}
\usepackage{amsmath,amssymb}
\usepackage{xcolor}
\usepackage{graphicx}
\usepackage{booktabs}
\usepackage{caption}
\usepackage{hyperref}
\usepackage{enumitem}
\usepackage{amsfonts,subfigure}
\usepackage{datetime}
\usepackage{microtype}
\usepackage{epsfig,color}
\usepackage{amsmath}
\usepackage[left,pagewise]{lineno} 
\usepackage[vlined,ruled]{algorithm2e}

\definecolor{orange}{rgb}{1.0,0.8,0.4}
\definecolor{red}{rgb}{1.0,0.0,0.0}
\definecolor{blue}{rgb}{0.0,0.0,1.0}
\definecolor{CadetBlue}{rgb}{0.5,0.0,0.9}
\definecolor{purple}{rgb}{0.7,0.3,0.8}

\begin{document}

\title{Space-Efficient Hidden Surface Removal}

\author
{    Frank Kammer\inst{1} 
\and Maarten L\"offler\inst{2}
\and Rodrigo I. Silveira\inst{3}
}

\institute
{    Institut f\"ur Informatik, Universit\"at Augsburg,
     \texttt{kammer@informatik.uni-augsburg.de}.
\and Department of Information and Computing Sciences, Utrecht University,
     \texttt{m.loffler@uu.nl}
\and Departament de Matem\`atiques, Universitat Polit\`ecnica de Catalunya,
     \texttt{rodrigo.silveira@upc.edu}
}

  \maketitle{}

  \begin{abstract}
 We propose a space-efficient algorithm for hidden surface removal that
 combines one of the fastest
 previous algorithms for that problem with techniques based on bit
 manipulation. 
Such techniques had been successfully used in other settings, for example to
reduce working space for several graph algorithms.  However, bit
manipulation is not usually employed in geometric algorithms because the
standard model of computation (the real RAM) does not support it.  For this
reason, we first revisit our model of computation to have a reasonable
theoretical framework.  Under this framework we show how the use of a bit
representation for the union of triangles, in combination with rank-select
data structures, allows us to implicitly compute the union of $n$ triangles
with roughly $O(1)$ bits per union boundary vertex. 
This results in an algorithm that uses at most as much space as the previous one, and depending on the input, can give a reduction of up to a factor $\Theta(\log n)$, while
 maintaining the running time.
 \end{abstract}

\pagestyle{plain}
\thispagestyle{plain}
\section {Introduction}

The search for algorithms that use as little storage as possible has received considerable attention in the last few years.
This is due in part to the increase in data volumes that currently need to be processed and analyzed, and also to the widespread use of devices that have limited memory, ranging from embedded systems to mobile phones.

The first papers on space-efficient algorithms considered
sorting~\cite{Bea91,PagR98} as well as
selection~\cite{ElmJKS14,Fre87,MunR96,RamR99}. 
More recently, space-efficient algorithms began to be studied for 
geometric and graph problems. 
The geometric problems studied include Delaunay triangulations and  
Voronoi diagrams~\cite{AsaMRW11,KorMRRSS15}, linear programming and convex hulls~\cite{ChaC07,DarE14}, visibility polygons~\cite{BarKLS14}, line segment intersections~\cite{KonA13}, and problems that can be solved by stack-based incremental algorithms (such as the construction of visibility polygons or polygon triangulations), among a few others~\cite{AsaBBKMRS14,AsaMRW11,BarKLSS13}.  
For a recent survey, we refer to
Korman~\cite{Korman16}.

The space-efficient algorithms for graph problems studied cover fundamental problems such as depth-first search, breadth-first search, computation of (strongly) connected components,
cutvertices and shortest paths~\cite{AsaIKKOOSTU14,ElmHK15,KamKL16}.

The setting where space-efficient algorithms are studied usually consists of a read-only input, a read-write working memory, and a write-only output memory. 
The general objective is to use as little working memory as possible.
However, the actual goals and techniques used for space-efficient algorithms in computational geometry and graph algorithms are different, mostly due to the different computation models assumed.
In computational geometry, the use of the \emph{real RAM} puts the focus on algorithms that use as few variables as possible.
In contrast, for space-efficient graph algorithms the model is often some variant of the \emph{word RAM}, and the goal is to minimize the size of the working
memory. To get a space bound that is independent from
the size of a word, the space consumption of those algorithms is
expediently measured in bits.
The use of bits in the representation of words allows us 
to use a powerful set of existing algorithms and data structures that work on bit representations, such as 
rank-select data structures for bit vectors~\cite{Cla96}
and choice dictionaries for sets~\cite{HagK16}.

\subsection{Computation model}

A large body of research in computational geometry focuses on the analysis
of the space and time requirements of algorithms that work on a {\em real
RAM}. 
The time complexity is measured in the total number of fundamental operations on
real numbers or integers, and the space complexity is the total
number of memory cells used.
In contrast, algorithms in other areas, such as graph algorithms, are often presented for variants of the \emph{word RAM}, in which space is measured in bits.
In this paper we are interested in applying some of the techniques
used successfully for graph
algorithms to geometric
problems, but at the same time, we want to keep the conceptual transparency
of the real RAM. We next briefly review these   models. 

\paragraph{Real RAM.}
A \emph{real random access machine}~\cite{blum1989,Preparata85} models an idealized computer that can manipulate arbitrary real numbers, and is the standard model of computation in computational geometry.
The model represents data as an infinite sequence of storage cells. 
These cells can be of two different types: cells that can store real numbers, or cells that can store integers.
The model supports standard operations on real numbers in constant time, including addition, multiplication, and elementary analytic functions such as taking roots, logarithms, trigonometric functions, etc.
The model also supports standard arithmetic operations on integers, and in addition, integers can be used to directly address memory cells.
In a sense, the model is a combination of a standard RAM (which we get by not using the real numbers), and a real-valued {\em pointer machine}~\cite{Knuth97} (which we get by never manipulating the
integers).

The true power of the real RAM lies in the combination of the two data types.
However, care must be taken: if we allow to freely convert real numbers to integers and vice versa, or indeed, if we can work with arbitrarily large integers at all, the model becomes unreasonably powerful and can solve PSPACE-complete problems in polynomial time~\cite{Schoenhage79}.
The literature is inconsistent in dealing with this issue, but often
a restricted floor function is (implicitly) assumed, that can convert, for instance, real numbers to their nearest integers in constant time only if the resulting integer is of polynomial size
w.r.t.\ the input.

\paragraph{Word RAM.}
A \emph{word RAM} is similar to a real RAM without support for real numbers and with a limited number of bits available to encode integers.
The word RAM represents data as a sequence of $w$-bit words, where it is usually assumed that 
$w=\Omega(\log n)$
where $n$ is the problem size.
Integers on a real RAM are usually treated as atomic, whereas the word RAM allows for powerful bit-manipulation tricks.
Data can be accessed arbitrarily, and standard
operations, such as Boolean operations 
(\texttt{and}, \texttt{xor}, \texttt{shl}, $\ldots$), addition, or
multiplication take constant time. 
One often assumes  that the input is read-only, there is read and write
access to the
working-space, and the output is write-only. Then, the space-consumption of an
algorithm is measured in the size of the required working-space.

There are many variants of
the word RAM, depending on precisely which instructions are 
supported in constant time. The general consensus seems
to be that any function in $\text{AC}^0$
is acceptable.\footnote{$\text{AC}^0$ is the 
class of all functions $f: \{0,1\}^* \rightarrow \{0,1\}^*$ that
can be computed by a family of circuits $(C_n)_{n \in \mathbb{N}}$ with the 
following properties: (i) each $C_n$ has $n$ inputs; (ii) there exist constants
$a,b$, such that $C_n$ has at most $an^b$ gates, for $n\in \mathbb{N}$; 
(iii) there is a constant $d$ such that for all $n$ the length of the longest
path from an input to an output in $C_n$ is at most $d$ (i.e., the
circuit family has bounded depth); (iv) each gate
has an arbitrary number of incoming edges (i.e., the \emph{fan-in} is 
unbounded).} However, it is always preferable to rely on a set of operations
as small, and as non-exotic, as possible.
Note that multiplication is not in $\text{AC}^0$~\cite{FurstSaSi84}. 
Nevertheless, it is usually
included in the word RAM instruction set~\cite{FredmanWi94}.

\paragraph{Bit manipulation in geometric algorithms.}

While the majority of geometric algorithms are analyzed on a pure real RAM, 
the advantage of bit manipulation 
and the fact that the word RAM more closely resembles real-life computers, has led to several researchers mixing the two models and treating the integers in a real RAM as
words~\cite{Bringmann2013,KrznaricLe98,Schrijvers:2013:DTW:2551692.2551706}.
When the model is handled carefully, this can lead to results that
can run on a real world computer within the same resource bounds and
that
are hard or impossible to obtain on a pure real RAM. 

However, these works only analyze the improved {\em time} complexity of such algorithms. The {\em space} complexity is harder to grasp---memory cells on a real RAM can store arbitrary numbers, while memory cells on a word RAM are restricted by their bits. The standard way to deal with this is to simply count all memory cells
equal---when using floating point arithmetic to approximate real numbers in real-life computers, this is not an unreasonable assumption.\footnote {Although, when real numbers are implemented using more sophisticated algebraic number types, their practical space consumption becomes much higher.} However, when the majority of memory cells used in an algorithm store integers, rather than real numbers, we may in principle be able to significantly improve the space complexity 
through bit manipulation. 

\paragraph{Model of choice.}

In this paper, we will adopt a 
real RAM with words for integers in the most pure sense.\footnote {We are not aware of a similar model of computation being explicitly described, despite the fact that it seems like a natural compromise between the word and real RAMs.} 
That is:

\begin {itemize} [noitemsep]
  \item real numbers are stored and manipulated in real-valued memory cells, as on a real RAM;
  \item integer and bits are stored and manipulated integer-valued memory cells, as on a word RAM;
  \item absolutely no conversion between the two kinds of cells is allowed
    and only integers can be used to address memory cells.
\end {itemize}

As it is often the case for algorithms on the word RAM, we assume that
the input is read-only and the output is write-only. In addition, we meassure the space consumption of our algorithms 
by the size of the required
working space.
Combining the word RAM and the real RAM is not new~\cite{bh-frram-98,ChanPa09};
however, 
most existing models are
not restrictive enough concerning the conversion between the two kinds of
cells, and one can exploit it to obtain algorithms that
using an unrealistic amount of working space
compared to what is possible on a real computer.

We
apply these techniques to one concrete geometric problem: the computation of visibility from one observer among a set of polyhedral obstacles. This problem is closely related to \emph{hidden surface removal}, a well-studied problem in computer graphics and computational geometry.
More precisely, we 
present 
a space-efficient algorithm for computing the viewshed of a point in a three-dimensional scene. 
We give a space-efficient implementation of Katz et al.'s
algorithm~\cite{DBLP:journals/comgeo/KatzOS92}
that computes the viewshed from a point in a three-dimensional scene composed of $n$ triangles.
The working space used by our algorithm consists of $O((U(n) + K) \log n)$ bits and $O(1)$ real numbers, where $K$ is a parameter that depends on the input, related to the complexity of unions of the input objects (a precise definition is given in Section~\ref{sec:algorithm}). 
In the worst case, $K$ can be $\Theta(U(n)\log n)$
and our new algorithm matches the working space used by Katz et al.'s algorithm.
However, we expect that in most practical situations, $K$ is closer to $\Theta(U(n))$, resulting in an improvement of a logarithmic factor.

Our main contribution is a concise representation of the union of triangles, together with a set of operations 
to manipulate them efficiently, which allows us to store intermediate results of the algorithm more efficiently. 
We choose Katz et al.'s algorithm
for two reasons.
Firstly, it is one of the fastest algorithms known for several types of scenes, including polyhedral terrains. This is relevant due to the many applications of this problem in geographic information systems.
Secondly, it is a conceptually simple algorithm, making it appropriate to try to apply our bit-based techniques.
Moreover, our technique works particularly 
well in the Katz et al.\ algorithm, because intermediate results are significantly larger than the final output.
We expect that the same \mbox{approach is applicable to more geometric
problems.}

We finally want to remark that there is a trivial algorithm to compute the viewshed of a
point that runs $O(n\ell)$ time and uses $O(\log (n+\ell))$ bits where $n$ is the number
of given polyhedral obstacles and $\ell$ denotes the total number vertices obtained
by intersecting all pairs of obstacles: Iterate over all pairs and test if
it is hidden by one obstacle. The viewshed consists of all boundaries of a
polyhedral obstacle that connects 2 not-hidden vertices.

\subsection {Hidden surface removal}

Given a set of objects in 3D and a viewing point $p$, a fundamental question
is to determine which parts of the objects are visible from $p$.  This is
sometimes called the \emph{viewshed} of $p$.  Equivalently, one may be
interested in determining the parts \emph{not} visible from $p$, which leads
to the \emph{hidden surface removal} problem.

Visibility problems of this type have been studied in computational geometry for a long time, due to the large number of applications that they have in computer graphics and geographic information systems (where the \emph{scene} usually consists of a polyhedral terrain).

It is well-known that in a scene with complexity $\Theta(n)$ (e.g., consisting of $n$ triangles), 
the viewshed of a viewpoint can have $\Theta(n^2)$ complexity, and can be computed in $O(n^2)$ time~\cite{McKenna87}.
Most practical algorithms are those that are \emph{output-sensitive}: their running time is proportional to the complexity of the viewshed, $k$.
The best running time for the most general case is achieved by the algorithm by Agarwal and Matousek~\cite{AgarwalM93}, although at the expense of a fairly complicated method.
Simpler but still efficient algorithms are known under the assumptions that a depth order among the 3D objects exists and can be computed efficiently (this is often, but not always, 
the case). For example, the algorithm by Goodrich~\cite{Goodrich92} runs in $O(n \log n + \ell + t)$ time, where $\ell$ is the number of intersecting pairs of line segments, and $t$ the number of intersections between scene polygons, in the projection plane (note that in our context, all polygons are triangles, thus $t=O(s)$).
The fastest algorithms under the depth-order assumption are the ones by Reif and Sen~\cite{rs-eoshsra-88} and Katz et al.~\cite{DBLP:journals/comgeo/KatzOS92}.
The former runs in time  $O((n + k) \log n \log \log n)$, while the second one has running time 
$O((U(n) + k)\log^2n)$  and uses $O(U(n)\log n)$ integer/real numbers, where
$U(n')$ is a  super-additive upper bound on the combinatorial complexity of the union of the projections of 
any $n'$ objects from the input ($U(n)$ is nearly-linear for many classes of objects, such as polyhedral terrains).

\subsection {Previous space-efficient algorithms}
As already mentioned, the first problems studied in the setting of space-efficient algorithms were sorting~\cite{Bea91,PagR98} and 
selection~\cite{ElmJKS14,Fre87,MunR96,RamR99}. 
Several 
researchers also considered
space-efficient algorithms for 
geometric problems. 
Asano et al.~\cite{AsaMRW11}
showed how to  
triangulate a planar point set and how to find a
Delaunay triangulation or a 
Voronoi diagram in $O(n^2)$ time with
$O(\log n)$ bits 
working space 
where $n$ denotes the number of given points.
Chan and Chen~\cite{ChaC07} presented
a randomized 
algorithm for linear programming that,
given an array of $n$ half-spaces in a constant number of dimensions,
computes the
lowest point in their intersection in $O(n)$
expected 
time and 
works with $O((\log n)^2)$ bits.
In addition, they described
a randomized algorithm for computing the convex hull of $n$
points sorted from left to right in the plane (i.e., in two dimensions) that
works with
 $O(n^\epsilon )$ bits and runs in $O(n/\epsilon)$ \mbox{expected time
for any fixed $\epsilon > 0$.}

Later, several papers with time-space trade-offs were published.
Darwish and Elmasry~\cite{DarE14} 
solved the convex-hull problem to optimality with an algorithm that
works with $\Theta(s)$ bits ($s\ge \log n$) and
runs in $O(n^2/s + n \log s)$~time.
An algorithm for computing a convex hull of a simple polygon was presented
by
Barba et al.~\cite{BarKLSS13}. They developed a general framework that
can be applied to incremental linear-time algorithms that, given $n$
objects, use a stack of 
size $O(n)$ and possibly $O(1)$ further variables.\footnote{In most of the previous work based on the real RAM, a \emph{variable} stands for either an integer or a real number.} The framework
allows
to reduce the space consumption of the algorithm to either $O(s)$ variables
($1 \le s \le \log n$)
at the price of an increased running time of
$O(n^2\log n/2^{s})$
or to $O(p \log n/\log p)$ variables for any $2 \le p \le n$ and time $O(n^{1+1/\log p})$. The framework can be used for
computing the
convex hull of a simple polygon as well as
a triangulation of a monotone
polygon, the shortest path between two given points inside a monotone
polygon,
and the
visibility profile of a point inside a simple polygon.
Moreover, the planar
convex-hull problem has been solved optimally with an algorithm that runs in
$O(n^2/s + n \lg s)$ time \cite{DarE14}.  Konagaya and Asano gave an
algorithm for reporting the line-segments intersections that runs in
$O((n^2/\sqrt{s}) \cdot \sqrt{\lg n}+ n \lg s + k)$ time \cite{KonA13},
where $k$ is the number of intersecting pairs.  Other papers that deal with
space-efficient geometric algorithms include
\cite{AsaBBKMRS14,AsaMRW11,BarKLSS13}.  Recently, Korman et al.~\cite{KorMRRSS15} gave space-efficient algorithms for triangulations and for
constructing Voronoi diagrams.

Furthermore,
Barba et al.~\cite{BarKLS14} described
an algorithm for computing the visibility of a 
simple polygon with $n$ vertices that works with 
only $O(1)$ variables (which can store integers or real numbers)
and
has a
running time of $O(nr)=O(n^2)$ where $r$ is the number of the so-called
{\em reflex vertices} of the polygon that are part of the output, and
Elmastry and Kammer~\cite{ElmK16} focused on space-efficient plane-sweep algorithms.

Elmasry et al.~\cite{ElmHK15}
presented several basic graph algorithms: They
showed that a depth-first search (DFS) can be carried out
in $O((n+m)\log n)$ time with 
$((\log_2\! 3)+\epsilon)n$ bits
for arbitrary fixed $\epsilon>0$.
A very similar result was found independently
by Asano et al.~\cite{AsaIKKOOSTU14}, who
 need $c n$ bits for an unspecified constant $c>2$,
or $\Theta(m n)$ time.  
Moreover, Elmasry et al.\ relaxed the space bound to $O(n)$ bits
at the price of an increased running time of
$O((n+m)\log\log n)$. 
In addition, they showed how to
run a DFS in reverse with only a modest       
penalty of $O(n\log\log n)$ additional        
bits.
Consequently, topological
sortings and strongly connected components
can be computed 
in linear time with $O(n\log\log n)$ bits.
Although the connected components of a given
undirected graph are usually computed by
means of DFS,                            
they observed that this bottleneck
can be avoided and showed how to output
the connected components in $O(n+m)$   
time with $O(n)$ bits
and how to compute a shortest path forest---and thus some variant of a
breadth-first search (BFS)---within the same resource bounds.

\subsection{Problem statement}

The input to our problem is a set $D$ of $n$ triangles in $\mathbb R^3$, and a {\em viewpoint} $p$. 
The input is given as a list of triples of points, where each point is in turn given as a triple of real
numbers.
We assume that there exists a compatible depth order on the triangles, as seen from $p$,
and we assume 
that the triangles 
are sorted in this order or that this order is
computable in negligible time and space.
Without loss of generality, we also assume that $n$ is a power of 2; otherwise,
add $\le n$ triangles inside one triangle. These new triangles do not modify
the solution.
We finally assume there are no three lines each extending a triangle edge 
\mbox{that intersect in one point.}

The output is a subdivision of each triangle into a {\em visible} and an {\em invisible} portion, where a point $q$ on a triangle is visible if the segment $pq$ does not intersect any other triangle of $D$. These visible portions are given as a list of polygons (possibly with holes). We denote their total complexity (that is, the total number of vertices of all these polygons together) by $k$.

\section{Hidden surface removal algorithm by Katz et al.}

\begin{figure}[h]
\centering
\includegraphics[scale=1]{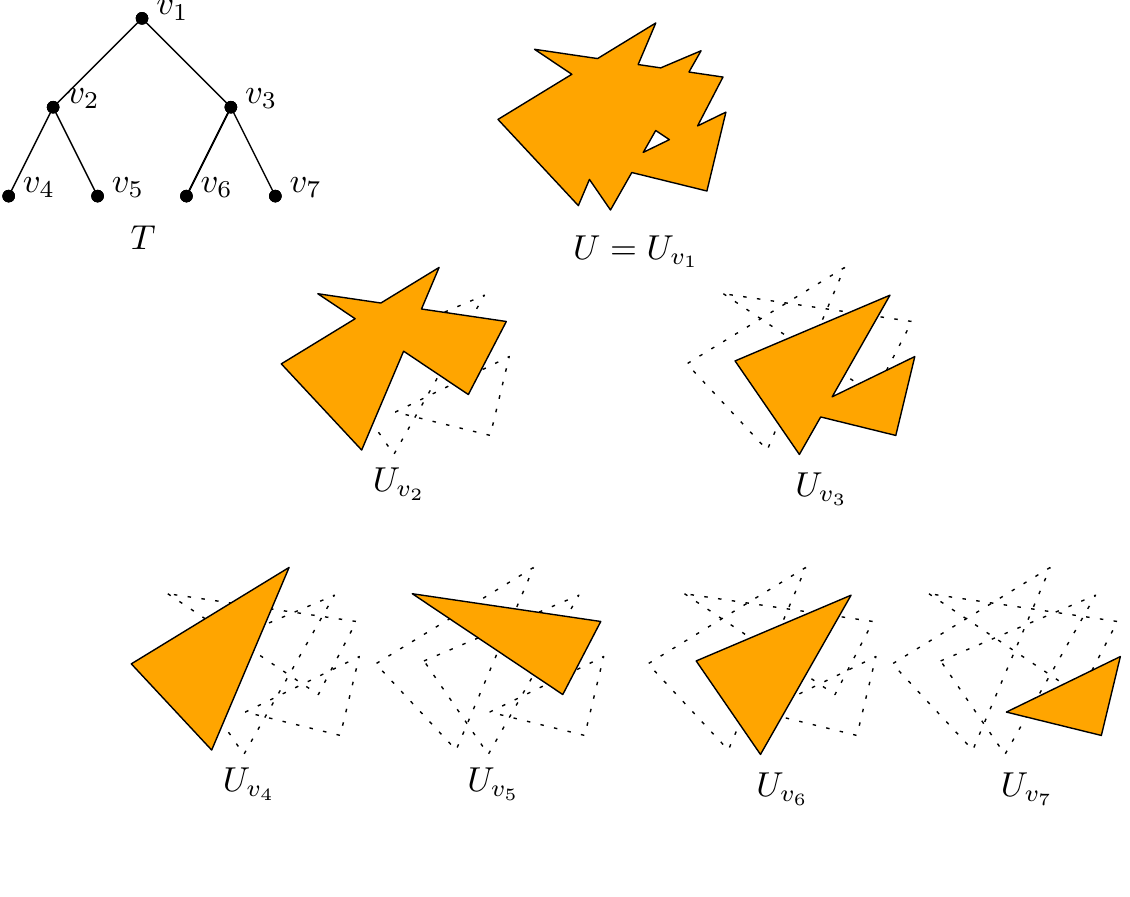}
\vspace{-4mm}
\caption{Partial unions $U_i$ associated to a tree $T$ for four triangles.}
\label{fig:Unions}
\end{figure}

We begin by describing the basic idea of the algorithm by Katz et al.~\cite{DBLP:journals/comgeo/KatzOS92}.
The input triangles are
stored in the leaves of a binary search tree $T$ in the given sorted order, with the nearest triangle in the 
rightmost
leave.
Each internal node $w \in T$ stores (1) the union $U_w$ of the projections
of the triangles in the subtree rooted at $w$ as well as 
(2) the visible portions $V_w$ of $U_w$ (i.e., visible with respect to \emph{all} input triangles).
Note that $U_w$ and $V_w$ are planar regions that may contain holes.
See Fig.~\ref{fig:Unions} for an example of the partial unions $U_i$.

The main task of the algorithm is to compute $V_w$ for all leaves $w$ since then gluing together all visible parts of triangles results in the output. 
To accomplish this, the algorithm first builds the partial unions $U_i$ in a bottom-up fashion, by computing, at each internal node, the union of the unions stored in both subtrees.
Once $U_{root}$ is built, the visible portions are produced by traversing $T$ recursively in preorder. At any time during the algorithm, only the visibility regions along one path are stored.
It follows that the space bottleneck of the algorithms comes 
from storing the $U_i$ in each of the nodes, that are required for the whole tree, adding up to $O(U(n) \log n)$ 
integer/real numbers. 

\section{Space-efficient union representation}
\label{sec:algorithm}

We present a method to build the tree of partial unions $U_i$ following the same recursive procedure as in~\cite{DBLP:journals/comgeo/KatzOS92}. The main difference will be in how each union is represented and stored.

\begin{figure}[h]
\centering
\includegraphics{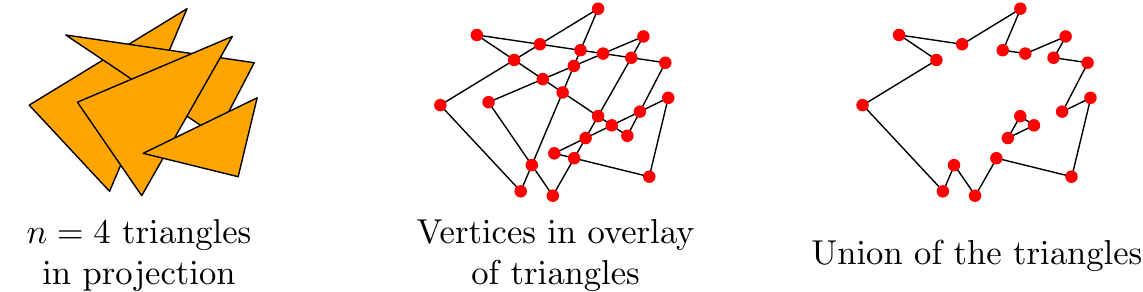} 
\caption{An example instance.}
\label{fig:example}
\vspace{4mm}

\includegraphics[scale=1]{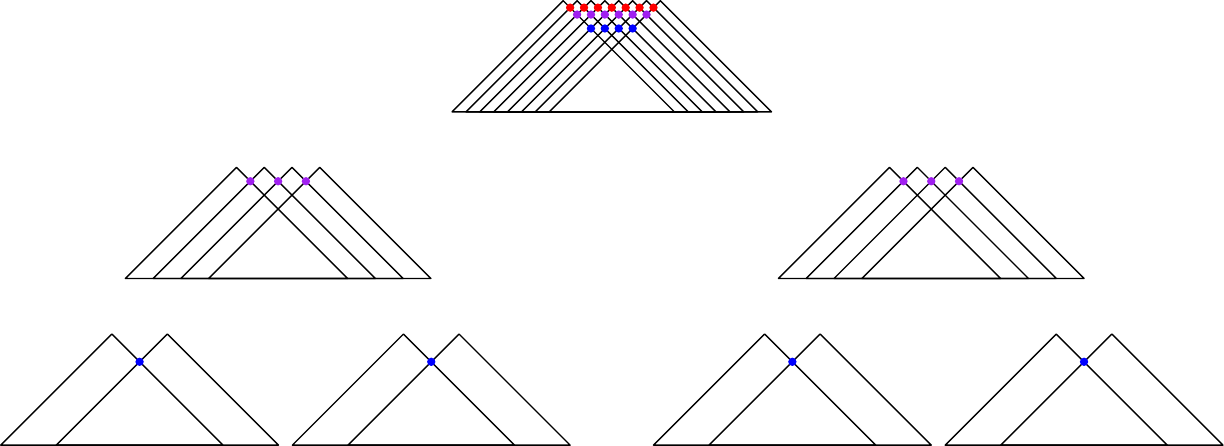}
\caption{Example with 8 triangles and a tree of partial unions such that in each node many union vertices appear.
Generalizing the construction to $n$ triangles results in $K = \Theta( U(n) \log n)$, which is  worst possible.}
\label{fig:fattriangles}
\end{figure}

The idea behind our method will be best illustrated with a running example. Consider the example in Fig.~\ref{fig:example}, showing the same four triangles as in Fig.~\ref{fig:Unions}.
The vertices of the union of $n$ triangles are a subset of those in their overlay (i.e., all pairwise edge intersections), or equivalently, vertices from the union of subsets of them.
However, as shown in the figure, not all of these vertices will show up in the union of the whole set.

The parameter $K$ is defined as the sum of the complexities of the partial unions $U_i$ over all levels of the tree. $K$ is $O(U(n)\log n)$ since the tree has height $O(\log n)$ and each level
of the tree corresponds to an instance of size $n$ with complexity $U(n)$. 
This bound is tight
since there are constructions for  which $K=\Theta(U(n)\log n)$
as shown in Figure~\ref{fig:fattriangles}.
 However, we point out that this situation occurs due to a combination of the actual geometry of the triangles with the way in which triangles have been grouped in the tree of partial unions. In practical situations, we expect such constructions to be uncommon.

Our method
represents partial unions by using bit vectors, so that each boundary vertex
of a partial union is encoded with $O(1)$ bits, instead of with $O(1)$ numbers.
Moreover, our algorithm processes the nodes of the tree 
with increasing heights of the nodes, starting from the leaves. 
We say that the nodes in the tree are processed {\em level by level},  
where level 1 consists of the leaves of the tree, level 2 is made of the parents of the leaves, and so on, until 
the root in level $\ell=\lceil \log_2 n \rceil+1$.

\begin{figure}[tb]
\centering
\includegraphics[angle=-90,scale=0.30]{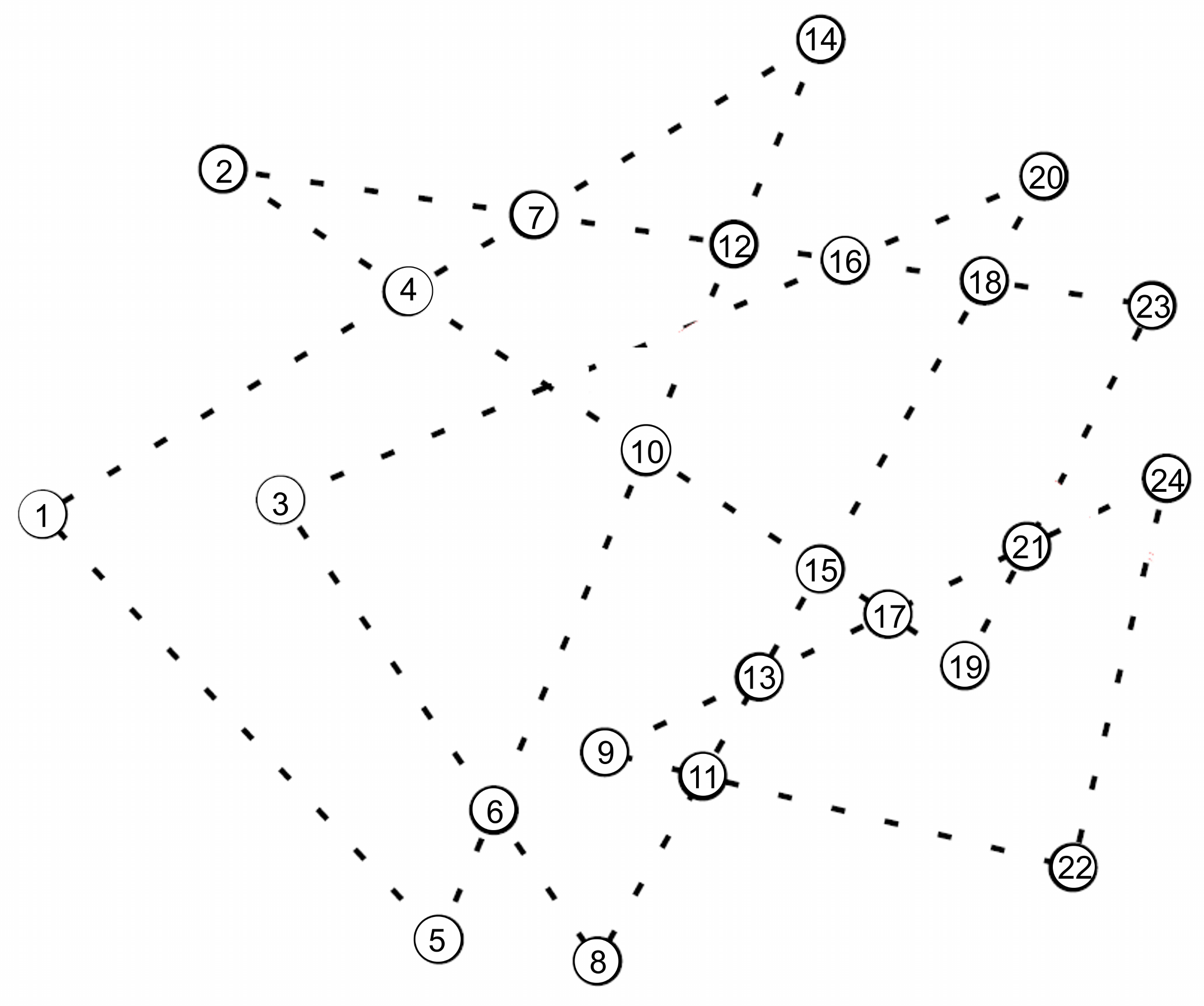}
\caption{Detail of triangles in Fig.~\ref{fig:example} showing vertex numbers for the vertices involved in the computation in Fig.~\ref{fig:Unions}.}
\label{fig:vertices}
\end{figure}

The three main ingredients of our representation will be a vector ${\mathcal C}_i$, a set of bit vectors associated with ${\mathcal C}_i$, denoted $B_{i,j}$, 
and a set of triangle bit vectors $A_w$. 
For $i\in \{1,\ldots,\ell\}$, let ${\mathcal C}_i$ be an array with 
all vertices that potentially can appear in $U_w$ for all nodes $w$ in levels
$1, \dots, i$, see Fig.~\ref{fig:vertices}.
It will be important to store ${\mathcal C}_{i}$ in a special way and to
store some extra information. 
Note that ${\mathcal C}_{i}$ and its extra
info will be stored only once and used for all nodes of the tree
that are in levels $1,\ldots, i$.

We assume that the triangles are numbered from $1$ to $n$. 
${\mathcal C}_i$ is stored in an array $C_i[1,\ldots,n]$ 
where each entry $C_i[k]$ is an array itself that
stores the vertices that are on the boundary of the $j$th triangle and that are used by some $U_w$ for a
node $w$ in some level $1 \dots i$.
Moreover, the vertices in  $C_i[k]$ are stored in the order found when walking along the boundary of a triangle $\triangle$ in clockwise direction.
Note that some vertices are the intersection of the boundaries of two triangles 
 $k'$ and $k''$, and thus they are stored in both arrays $C_i[k']$ and $C_i[k'']$.
We also store {\em cross pointers} between these two entries---a concept
 introduced for undirected graphs~\cite{ElmHK15}.
For each triangle $k$, we also store the number $p_i$ of vertices in $C_i[k]$ as well as the prefix sums
$P_x=\sum_{y=1}^x p_y$.  

All the arrays $C_i[1],C_i[2],C_i[3],\ldots$ are stored 
in consecutive order and we consider it as a global array, which we identify
with the name ${\mathcal C}_i$. 
Thus, each vertex in the set ${\mathcal C}_i$ has an {\em absolute position} 
 $p\in \{1,\ldots,|{\mathcal C}_i|\}$ in the array ${\mathcal C}_i$.
Using the prefix sums we can translate between the $j$th vertex of the $k$th triangle and 
its absolute position $p$ in ${\mathcal C}_i$. 
See Table~\ref{table:vectors} for the vectors that correspond to our example.

\begin{table}[h]
\scalebox{0.83}[0.83]{
\begin{tabular}{ll}
${{\mathcal C}_1}$&:$\!$\small{\texttt{ 1,14, 5| 2,23,19| 3,20, 8| 9,24,22}} \\
   ${B_{1,1}}\!\!$&:$\!$\small{\texttt{ 1, 1, 1| 1, 1, 1| 1, 1, 1| 1, 1, 1}}\vspace{2mm}\\
${{\mathcal C}_2}$&:$\!$\small{\texttt{ 1, 4, 7,14,12,10, 5| 2, 7,12,23,19,10, 4| 3,20,13,11, 8| 9,13,24,22,11}}\\
   ${B_{2,1}}\!\!$&:$\!$\small{\texttt{ 1, 0, 0, 1, 0, 0, 1| 1, 0, 0, 1, 1, 0, 0| 1, 1, 0, 0, 1| 1, 0, 1, 1, 0}}\\
   ${B_{2,2}}\!\!$&:$\!$\small{\texttt{ 1, 1, 1, 1, 1, 1, 1| 1, 1, 1, 1, 1, 1, 1| 1, 1, 1, 1, 1| 0, 1, 1, 1, 1}}\vspace{2mm}\\
${{\mathcal C}_3}$&:$\!$\small{\texttt{ 1, 4, 7,14,12,10, 6, 5| 2, 7,12,16,18,23,21,19,17,15,10, 4| 3,16,20,18,15,13,11, 8, 6| 9,13,17,21,24,22,11}}\\
   ${B_{3,1}}\!\!$&:$\!$\small{\texttt{ 1, 0, 0, 1, 0, 0, 0, 1| 1, 0, 0, 0, 0, 1, 0, 1, 0, 0, 0, 0| 1, 0, 1, 0, 0, 0, 0, 1, 0| 1, 0, 0, 0, 1, 1, 0}}\\
   ${B_{3,2}}\!\!$&:$\!$\small{\texttt{ 1, 1, 1, 1, 1, 1, 0, 1| 1, 1, 1, 0, 0, 1, 0, 1, 0, 0, 1, 1| 1, 0, 1, 0, 0, 1, 1, 1, 0| 0, 1, 0, 0, 1, 1, 1}}\\
   ${B_{3,3}}\!\!$&:$\!$\small{\texttt{ 1, 1, 1, 1, 1, 0, 1, 1| 1, 1, 1, 1, 1, 1, 1, 0, 1, 1, 0, 1| 0, 1, 1, 1, 1, 1, 1, 1, 1| 0, 1, 1, 1, 1, 1, 1}}\\
\end{tabular}
}%
\vspace{2mm}

\caption{Vectors  ${{\mathcal C}_i}$ and $B_{i,j}$ for our example.}
\label{table:vectors}
\end{table}

Based on ${\mathcal C}_i$, we can define bit vectors $B_{i,j}$.
For each level $j \le i$, $B_{i,j}$ is a bit vector with the same size as ${\mathcal C}_i$, following the same triangle structure as ${\mathcal C}_i$: each group of consecutive entries represents one of the input triangles. 
$B_{i,j}$ is defined as follows:  $B_{i,j}[k]=1$ exactly
when the vertex ${\mathcal C}_i[k]$ appears in the partial union of level $j$ that includes the corresponding triangle.
This means that $U_w$ and $U_{w'}$ with $w\neq w'$ are
stored in the same bit vector $B_{i,j}$ whenever $w$ and $w'$ are in the same level
$j$.
This is possible since 
the intersection points are pairwise disjoint in each level
and thus
each vertex is part of at most one set $U_w$ or
$U_{w'}$. 
We refer again to Table~\ref{table:vectors} for an example.

Finally, we
store for each node $w$ a bit vector $A_w$ over the triangles
that are identified with descendants of $w$ as follows: there is a $1$ for
a triangle exactly when
the boundary of the triangle \mbox{is part of the boundary of $U_w$.}

With these data structures in place, consider now the computation of  $U_w$ for all nodes $w$ in level
$i+1$, given the bit vectors $B_{i,j}$ over ${\mathcal C}_{i}$ for each level $j< i+1$. 
Assume for now that based on the bit vectors $B_{i,j}$ we can determine $U_{w'}$ for
all nodes $w'$ in a level $j< i+1$ (we defer the details of this to the next section).
Then we can compute $U_w$ for all nodes $w$
in level $i+1$. In particular, we can determine the set ${\mathcal C}_{i+1}$ as the union of
${\mathcal C}_{i}$ and the new vertices found 
on the boundary $U_w$ of some node $w$ in level $i+1$, which can be computed using 
the standard intersection algorithm by Bentley and Ottmann~\cite{BenO79}.
However, instead of storing intersection points using real-valued memory
cells, we store them implicitly, by storing the indices of the two segments of the input that generate the intersection point. This allows us to store all necessary information for a vertex using $O(\log n)$ bits.

Note also that for each $U_w$ of the nodes $w$ in the tree,
we store a pointer to a vertex in ${\mathcal C}_i$ that can be used 
as start vertex to traverse the boundary components of $U_w$. 
In our example, the pointer could point to the following vertices:
$v_1: 1,13$ (we store two pointers since $U_{v_1}$ has an outer boundary and a hole)
$v_2: 1$, $v_3: 3$, $v_4: 1$, $v_5: 2$, $v_6: 8$, $v_7: 22$. 

It is important to note that at any time during the algorithm, we only
need to maintain $C_i$ and the $B_{i,j}$ vectors for the previous and current level.

\subsection{Reconstructing lower-level unions with rank-select data structures}

In order to determine $U_{w'}$ for nodes $w'$ in level $i+1$, we need to know $U_w$ for nodes $w$ in levels 1 to $i$.
In this section we describe how to use the bit vector $B_{i,j}$ over ${\mathcal C}_i$ of level $j\le i$
to reconstruct $U_w$ for a node $w$ in level $j$.
The key ingredient is to build a 
rank-select data structure on each bit vector $B_{i,j}$ and each bit vector
$A_w$.

\paragraph{Rank-select data structures.}
A rank-select data structure for a bit sequence 
$B=(b_1,\ldots,b_N)$ is a data structure that
supports two types of queries:
$\mathrm{rank}_B(j)$ ($j\in\{1,\ldots,N\}$),
which returns $\sum_{i=1}^j b_i$; and
$\mathrm{select}_B(k)$ ($k\in\{1,\ldots,\sum_{i=1}^N b_i\}$),
which returns the smallest $j\in\{1,\ldots,N\}$
with $\mathrm{rank}_B(j)=k$.
It is well-known that rank-select structures
for bit sequences of length~$N$ that support
rank and select queries in constant
time and occupy $O(N)$ bits of space can
be constructed in $O(N)$ time \cite{Cla96}.
All rank-select data structures
introduced below are of this type.

\paragraph{Computing $U_w$.}
To compute a set
$U_w$ for a node $w$ in level $j$, we proceed as follows: Color all triangles
white (more precisely, always have a color array where all triangles are white,
then use it, and at the end of the usage, undo the recoloring). 
Using the rank select structure over $A_w$
we determine a white triangle $k'$ that has some common boundary with
$U_w$, then we use the rank-select structure on $B_{i,j}$ to find a first
vertex that is on the boundary of $U_w$ and of $k'$, 
and can translate the absolute position of the vertex to a
relative position in $C_i[k']$.
We next start an iteration to find the rest
of the closed curve around
the boundary of $U_w$. We always 
know
a vertex part of the boundary of a triangle $k'$; 
and this vertex is either the corner of a triangle or an intersection point of $k'$
with another triangle.
Making use of the rank-select structures we can skip over the corners and
assume without loss of generality that the current vertex is an intersection point of $k'$
with another triangle $k''$.
More exactly, we know the position of the vertex
in $C_i[k']$ where we can 
follow a cross pointer to the position of the same vertex in $C_i[k'']$. Using the prefix sums we get the
absolute position of the vertex in ${\mathcal C}_i$. The
rank-select structure allows us to find the next vertex on the closed curve, which is
w.l.o.g.\ a
vertex of an intersection
point between the triangle $k''$ and another triangle $k'''$. Following
again a cross
pointer we can now jump to that vertex in $C_i[k''']$.
 Whenever we extend the boundary
 by some vertex, 
we test if all vertices of the current triangle are now part of the boundary (using a
 separate counter
 for each triangle). If so, 
we color the triangle black. 
After we have found a closed curve, 
if there are still white triangles in $A_w$, then the boundary of
 $U_w$ has holes, so we we rerun the procedure.

We illustrate this in our example by showing how to 
reconstruct $U_{v_3}$ using ${{\mathcal C}_3}$ and ${B_{3,2}}$. We start
at vertex 3 in ${{\mathcal C}_3}$. Using the rank-select structures, we find
the vertex after 3 that also has a 1 in ${B_{3,2}}$, which is vertex 20.
Since 20 is a vertex of a triangle (20 has no cross pointer), we are looking
for the next 1 in ${B_{3,2}}$, which is vertex 13.
Then we use the cross pointer to jump to the other 13 in ${{\mathcal C}_3}$.
Again, we are looking for the next 1s in ${B_{3,2}}$ and find so
24 and subsequently 22 in ${{\mathcal C}_3}$, which are both corners of a
triangle. So we continue and
find 11. Using again a
cross pointer, we jump to the other 11 in ${{\mathcal C}_3}$, search for the next 1 in
${B_{3,2}}$
and find 8. Searching for the next one in ${B_{3,2}}$, we find 3 since we
have to consider each part as cyclic. Since 3 is the vertex where we begun,
we are done.

It remains to analyze the
required working space. While processing the nodes in level $i$, we
store vertex numbers in ${{\mathcal C}_{i^*}}$ and suitable cross pointers
for levels $i^*\in \{i-1,i\}$.
Since each vertex number and each cross pointer can be stored with $O(\log n)$ bits, 
and ${{\mathcal C}_{i^*}}$ cannot have more vertices that those that appear over the partial unions in the tree, 
${{\mathcal C}_{i^*}}$ can be stored with $(O(U(n)+K)\log n)$ bits, for the whole tree. In addition, we have
$O(\log n)$ bit vectors
${B_{i^*,1}},\ldots,{B_{i^*,i^*}}$ of $O(U(n)+K)$ bits each. In total, the
algorithm uses 
$O((U(n)+K)\log n)$ bits.

\begin{theorem}
There is an algorithm 
that
reports
the union of a set of $n$ non-intersecting
triangles in 3D in time $O((U(n) +
k)\log^2n)$ by using $O((U(n)+K)\log n)$ bits of working space 
and $O(1)$ 
real numbers, 
where
$U(n')$ is  a super-additive bound on the maximal
complexity  of the union of any $n’$ triangles from the family under consideration, $k$ is the         
complexity of the output, and $K$ is the sum of the complexities of the partial unions over all levels of the recursion tree used by the algorithm.
\end{theorem}

\section{Application to the algorithm by Katz et al.}

As mentioned before, the space bottleneck in the algorithm by Katz et al.\ is the storage of the partial unions $U_w$.
Therefore we can directly apply our technique, replacing the representation of the partial union boundaries by our bit-based representation, automatically reducing the storage used in terms of bits.

The only detail remaining is how to store $V_w$. In contrast to the sets
$U_w$, we do not have the property that the sets $V_w$ and $V_{w'}$ of one
level have disjoint vertices. Thus, we use one bit vector over $U$ for each
such set. Concerning the space consumption this is no problem since we have to store such a bit vector
only for the nodes that are part of a root-to-leaf path in the tree.

\begin{theorem}
Consider a set of $n$ non-intersecting triangles in space and a viewing point
 $p$, such that there exists a known depth ordering of the objects with
 respect to $p$, and such that the union of the projections of any $n'$ of
 the objects on a viewing plane has complexity $U(n')$, where $U(n')$ is
 super-additive.
 Then the visibility map from $p$ can be reported
  with 
 $O((U(n) + k)\log^2n)$ time, using $O((U(n)+K)\log n)$ bits of working
 space and
 $O(1)$ 
 real numbers,
 where $k$ is the
complexity of the visibility map,  and $K$ is the sum of the complexities of the partial unions over all levels of the recursion tree used by the algorithm.
\end{theorem}

\section{Conclusion}

We have shown that techniques previously used for graph algorithms can also be applied to
geometric
problems.
In line with recent results for graph
algorithms~\cite{AsaIKKOOSTU14,ElmHK15,KamKL16}, 
the space consumption to compute the viewshed of a point in a three-dimensional scene 
can be reduced by a factor of $\Theta(\log n)$ while maintaining
the running time. 
However, the space used ultimately depends on the complexities of the intermediate unions along the recursion tree, represented by the parameter $K$, which sometimes can be $\Theta(U(n) \log n)$. 
It may be possible to reduce the dependency on the intermediate unions by storing only the union vertices in each level that contribute to the current union, and not the rest. 
Exploring this direction further is an interesting direction for further research.

\section*{Acknowledgements}
M.L. is partially supported by the Netherlands Organisation for Scientific Research (NWO) through project 614.001.504.
R.I.S was partially supported by projects MTM2015-63791-R (MINECO/FEDER) and Gen.\ Cat.\ DGR2014SGR46, and by MINECO through the Ram{\'o}n y Cajal program.
We thank Wolfgang Mulzer for point out a mistake in a previous version of this paper.
\bibliographystyle{plain}
\bibliography{refs,main}

\end{document}